\documentclass[preprint,showpacs,preprintnumbers,amssymb,aps,nofootinbib]{revtex4}
\usepackage{amsfonts,color,graphicx,epsfig,graphicx,amsmath,multirow,slashed}

\begin{document}

\title{Impact of $Z \to \eta_{c,b}+g+g$ on the inclusive $\eta_{c,b}$ meson yield in $Z$-boson decay}
\author{Zhan Sun}
\email{zhansun@cqu.edu.cn}
\author{Xuan Luo}
\author{Ying-Zhao Jiang}
\affiliation{
Department of Physics, Guizhou Minzu University, Guiyang 550025, People's Republic of China.
} 

\date{\today}

\begin{abstract}
In this paper, we carry out the next-to-leading-order QCD corrections to $Z \to \eta_Q+g+g~(Q=c,b)$ (labeled as $gg$) through the color-singlet (CS) state of $Q\bar{Q}[^1S_0^{[1]}]$, with the aim of assessing the impact of this process on $Z$ bosons decaying into inclusive $\eta_Q$. We find that the QCD corrections to the $gg$ process can notably enhance its leading-order results, especially for the $\eta_c$ case, which would then greatly increase the existing predictions of $\Gamma_{Z \to \eta_Q+X}$ given by the CS-dominant process $Z \to \eta_{Q}[^1S_0^{[1]}]+Q+\bar{Q}$. Moreover, with these significant QCD corrections, the $gg$ process would exert crucial influence on the CS-predicted $\eta_Q$ energy distributions. In conclusion, in the CS studies of $Z \to \eta_Q+X$, besides $Z \to \eta_{Q}[^1S_0^{[1]}]+Q+\bar{Q}$, $Z \to \eta_Q[^1S_0^{[1]}]+g+g$ can provide phenomenologically indispensable contributions as well.

\end{abstract}

\maketitle

\section{Introduction}

Due to experimental reconstruction difficulties,\footnote{$\eta_c$ is always established by its decay into multiple hadrons, such as $p\bar{p}$, which is more difficult than the $J/\psi$ detection.} the observation of the $\eta_c$ meson is scant compared to that of $J/\psi$. For example, HERA, LEP II, and $B$ factories have accumulated copious $J/\psi$ yield data, but they have not yet detected any evident event of inclusive $\eta_c$ production. In 2014, the LHC (LHCb group), which runs with a large center-of-mass proton-proton collision energy and a high luminosity, achieved the first measurement of inclusive $\eta_c$ yield \cite{LHCb:2014oii}. Compared to the theoretical results \cite{etac1,etac2,etac3,etac4,etac5,etac6,etac7,etac8,etac9,etac10,
etac11,etac12}, the measured cross sections seem to almost be saturated by the color-singlet (CS) predictions alone, leaving very limited room for the color-octet contributions and thus posing a serious challenge to the nonrelativistic QCD (NRQCD) factorization \cite{NRQCD}; however, Refs. \cite{etac4,etac5} point out that NRQCD is still valid in describing the LHCb data. Note that, there are large uncertainties in the LHCb released data \cite{LHCb:2014oii}. Therefore, more studies of inclusive $\eta_c$ yield in other processes and experiments with better precision are required to further assess the validity of NRQCD in $\eta_c$ production.

Heavy-quarkonium production in $Z$-boson decay, which has triggered extensive studies \cite{z decay 2,z decay 3,z decay 4,z decay 5,z decay 6,z decay 7,z decay 8,z decay 9,z decay 10,z decay 11,z decay 12,z decay 13,z decay 14,z decay 15,z decay 16,z decay 17,z decay 18,z decay 19,z decay 20,z decay 21,z decay 22,z decay 23,z decay 24,z decay 25,z decay 26,z decay 27,z decay 28}, provides a good chance for studying the $\eta_c$ production mechanism. At the LHC, a large number of $Z$ events ($\sim10^{9}$/year \cite{z decay 20}) can be generated in one running year, with which the study of $Z$ decaying into heavy quarkonium has been an increasingly important area \cite{LHC_exp1,LHC_exp2,LHC_exp3}. Furthermore, the upgrades of HE(L)-LHC will give birth to a higher collision energy (luminosity), largely improving the accumulated $Z$ yield events. In addition, the proposed future $e^+e^-$ collider, CEPC \cite{CEPC}, equipped with a $``\textrm{clean}"$ background and an enormous $Z$ production events ($\sim10^{12}$/year), would also be beneficial for hunting $Z$ decaying into inclusive $\eta_c$. From these perspectives, a precise measurement of $Z \to \eta_c+X$ looks promising, and the theoretical study of this process through the CS mechanism could help to explore whether the compatibility of the CS predictions with future measurements still holds.  

In $Z \to \eta_c+X$, there exist two CS processes contributing at leading-order (LO) accuracy in $\alpha_s$, i.e., $Z \to \eta_c[^1S_0^{[1]}]+c+\bar{c}$ (labeled as $c\bar{c}$) and $Z \to \eta_c[^1S_0^{[1]}]+g+g$ (labeled as $gg$). We can learn from Refs. \cite{z decay 4,z decay 5} that the $c\bar{c}$ process plays a leading role in the CS LO predictions because of the $c$-quark fragmentation; owing to the suppression of $\frac{m_c^2}{m_Z^2}$ \cite{z decay 5}, the $gg$ process contributes just slightly at LO (less than $5\%$ of $\Gamma_{c\bar{c}}$). However, considering the advent of the gluon-fragmentation structures in the next-to-leading-order (NLO) calculations of $Z \to \eta_c+g+g$, i.e., $Z \to q+\bar{q}+g^{*}$;$g^{*} \to \eta_c+g$ ($q=u,d,s$) and the loop-induced process $Z \to g+g^{*}$;$g^{*} \to \eta_c+g$, the uncalculated QCD corrections to the $gg$ process are expected to provide considerable contributions. In addition, the $\eta_c$ energy distributions in the $gg$ and $c\bar{c}$ processes may thoroughly be different. The $gg$ process, together with the QCD corrections, are strongly suppressed by the factor $\frac{M_{\eta_c}^2}{E_{\eta_c}^2}$ for large-$z$ \cite{z decay 13,Kuhn:1981jn,Kuhn:1981jy}, and thereby the $z$ value corresponding to the largest $\frac{d\Gamma}{dz}$ should be small; regarding the $c\bar{c}$ process, as a result of the $c$-quark fragmentation, the dominant contributions exist in the large $z$ region \cite{z decay 5}. In view of these points, $Z \to \eta_c[^1S_0^{[1]}]+g+g$ would be phenomenologically crucial for the inclusive $\eta_c$ yield in $Z$ decay.

In contrast with $\eta_c$, the larger mass of $\eta_b$ would result in a smaller typical coupling constant and relative velocity ($v$) between the constituent $b\bar{b}$ quarks, subsequently leading to better convergent results over the expansion in $\alpha_s$ and $v$. On the experimental side, however, $\eta_b$ has so far been observed only in $e^+e^-$ annihilation \cite{etab1,etab2,etab3,etab4}. Taken together, in this article we will carry out the first NLO QCD corrections to $Z \to \eta_c(\eta_b)[^1S_0^{[1]}]+g+g$, so as to provide a deeper insight into the $\eta_c(\eta_b)$ production mechanism.

The rest of the paper is organized as follows: In Sec. II, we give a description of the calculation formalism. In Sec. III, the phenomenological results and discussions are presented. Section IV is reserved as a summary.

\section{Calculation Formalism}

Within the NRQCD framework \cite{NRQCD,Petrelli:1997ge}, the decay width of $Z \to \eta_Q+X~ (Q=c,b)$ can be factorized as
\begin{eqnarray}
\Gamma=\hat{\Gamma}_{Z \to Q\bar{Q}[n]+X}\langle \mathcal O ^{\eta_Q}(n)\rangle,\label{fac exp}
\end{eqnarray}
where $\hat{\Gamma}$ are the perturbative calculable short distance coefficients (SDCs), representing the inclusive production of a configuration of the $Q\bar{Q}[n]$ intermediate state. The universal nonperturbative long-distance matrix element $\langle \mathcal O ^{\eta_Q}(n)\rangle$ stands for the probability of $Q\bar{Q}[n]$ into $\eta_Q$. In this paper, we focus only on the CS contributions, and accordingly $n$ takes on $^1S_0^{[1]}$. The LO process of $Z \to Q\bar{Q}[^1S_0^{[1]}]+Q+\bar{Q}$, which is introduced as a comparison and which is free of divergence, has been calculated in Ref. \cite{z decay 4}; in the following, we only describe the calculation formalism of $Z \to Q\bar{Q}[^1S_0^{[1]}]+g+g$ up to the NLO QCD accuracy.

\subsection{LO}

The LO SDCs can be expressed as
\begin{eqnarray}
\hat{\Gamma}_{\textrm{LO}}=\int|\mathcal{M}|^2 d\Pi_3,
\end{eqnarray}
where $|\mathcal{M}|^2$ is the squared amplitude, and $d\Pi_3$ is the standard three-body phase space.

\begin{figure*}
\includegraphics[width=0.46\textwidth]{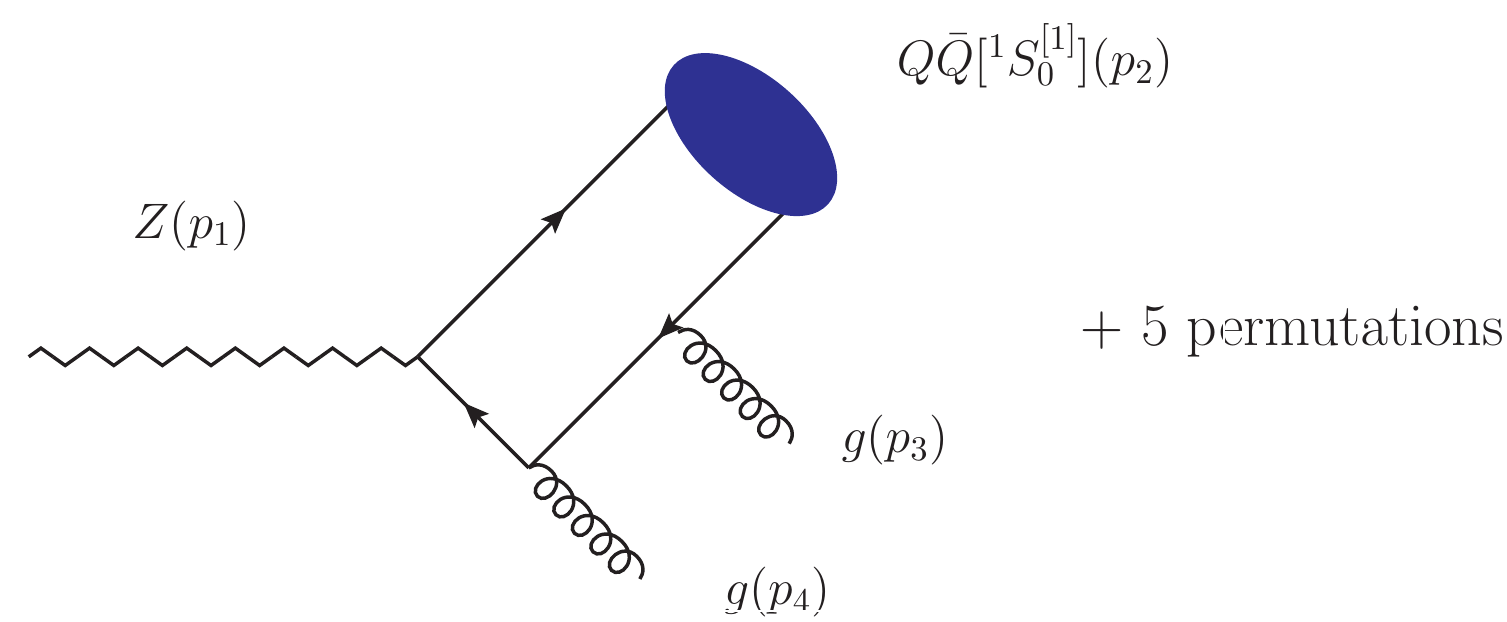}
\caption{\label{fig:LO}
Typical LO Feynman diagrams of $Z \to Q\bar{Q}[^1S_0^{[1]}]+g+g$ ($Q=c,b$).}
\end{figure*}

According to Fig. \ref{fig:LO}, $\mathcal{M}_{1}$ can be written as
\begin{eqnarray}
\mathcal{M}_{1}=\kappa \times \textrm{Tr} \left[\epsilon\!\!\!\slash(p_1)(\xi_1 P_L+\xi_2 P_R)\frac{-p\!\!\!\slash_{22}-p\!\!\!\slash_{3}-p\!\!\!\slash_{4}+m_Q}{(p_{22}+p_3+p_4)^2-m_Q^2} \epsilon\!\!\!\slash(p_4) \frac{-p\!\!\!\slash_{22}-p\!\!\!\slash_{3}+m_Q}{(p_{22}+p_3)^2-m_Q^2} \epsilon\!\!\!\slash(p_3) \Pi^{0}_{Q\bar{Q}}(p_2) \right], \nonumber \\
\end{eqnarray}
where $\kappa=\mathcal{C} \frac{e g_s^2}{4 \sin\theta_{\textrm{w}} \cos\theta_{\textrm{w}}}$, with $\mathcal{C}$ being the color factor. $\epsilon(p_{1})$ and $\epsilon(p_{3(4)})$ are the polarization vectors of the initial $Z$ boson and the final-state gluons, respectively. $P_L=(1-\gamma^{5})/2$ and $P_R=(1+\gamma^{5})/2$; $\xi_1=2-\frac{8}{3}\sin^2\theta_{w}$ and $\xi_2=-\frac{8}{3}\sin^2\theta_{w}$ for the $Z_{c\bar{c}}$ vertex, while $\xi_1=2-\frac{4}{3}\sin^2\theta_{w}$ and $\xi_2=-\frac{4}{3}\sin^2\theta_{w}$ for the $Z_{b\bar{b}}$ vertex.

The momenta of the constituent quarks follow as
\begin{eqnarray}
p_{21}=\frac{m_Q}{M_{Q\bar{Q}}}p_2+q~~\textrm{and}~~p_{22}=\frac{m_Q}{M_{Q\bar{Q}}}p_2-q,
\end{eqnarray}
where $m_{Q(\bar{Q})}=M_{Q\bar{Q}}/2$ is implicitly adopted to ensure the gauge invariance of the hard scattering amplitude; $q( \simeq 0)$ is the relative momentum between the two constituent heavy quarks inside the quarkonium.

The covariant form of the projector $\Pi^{0}_{Q\bar{Q}}$ reads 
\begin{eqnarray}
\Pi^{0}_{Q\bar{Q}}(p_2)=\frac{1}{\sqrt{8m_Q^3}}(p_{22}-m_{\bar{Q}})\gamma^{5}(p_{21}+m_{Q}).
\end{eqnarray}

In a similar way, the amplitudes $\mathcal{M}_{2},...,\mathcal{M}_{6}$ can be derived by permutations. By squaring the sum of all six amplitudes and summing over the polarization vectors of the $Z$ boson and the two final gluons, we finally obtain the squared amplitude $|\mathcal{M}|^{2}$.  

\subsection{NLO}

Up to NLO in $\alpha_s$, the SDCs comprise three contributing components,
\begin{eqnarray}
\hat{\Gamma}_{\textrm{NLO}}=\hat{\Gamma}_{\textrm{Born}}+\hat{\Gamma}_{\textrm{Virtual}}+\hat{\Gamma}_{\textrm{Real}},
\end{eqnarray}
where $\hat{\Gamma}_{\textrm{Born}}$ refers to the tree-level process and $\hat{\Gamma}_{\textrm{Virtual}(\textrm{Real})}$ is the virtual (real) correction.

\subsubsection{Virtual corrections}

\begin{figure*}
\includegraphics[width=0.95\textwidth]{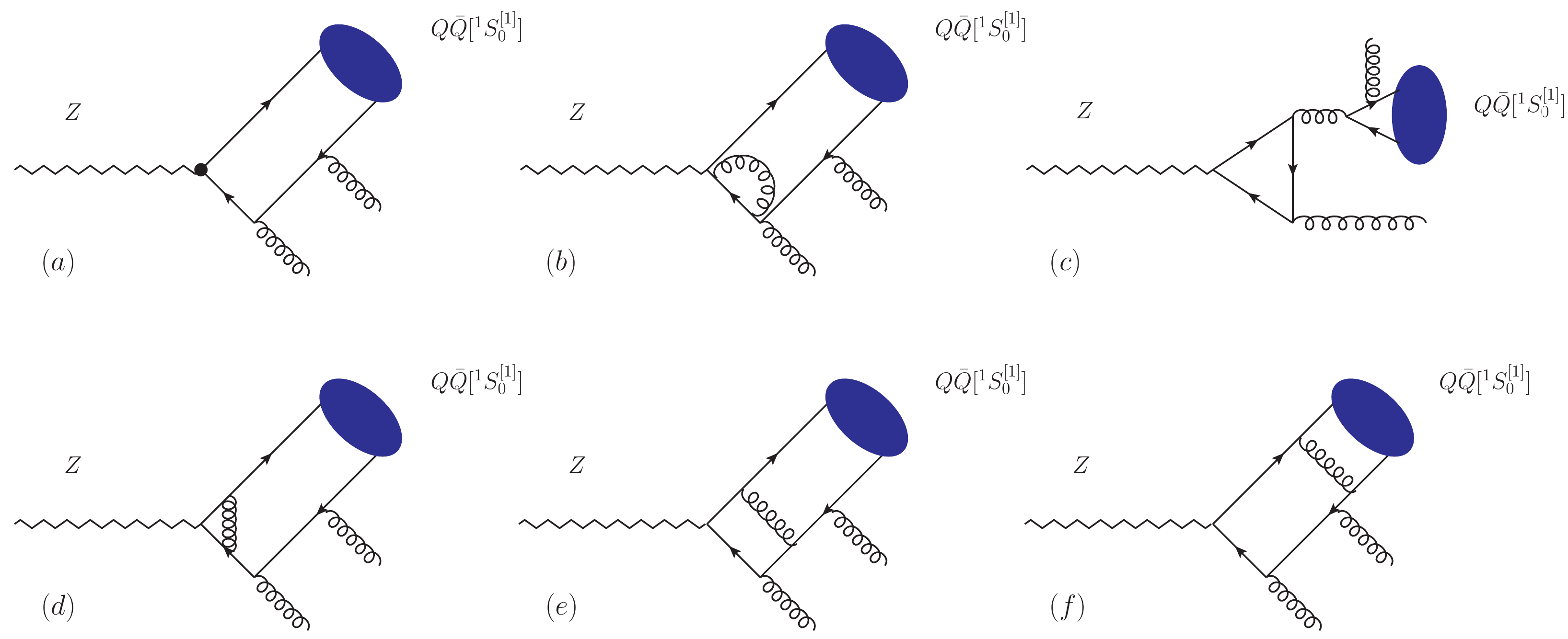}
\caption{\label{fig:Virtual}
Representative Feynman diagrams of the virtual corrections to $Z \to Q\bar{Q}[^1S_0^{[1]}]+g+g$ ($Q=c,b$).}
\end{figure*}

The virtual corrections are composed of the contributions of the one-loop ($\hat{\Gamma}_{\textrm{Loop}}$) and counterterm ($\hat{\Gamma}_{\textrm{CT}}$) diagrams, as representatively shown in Fig.  \ref{fig:Virtual}. $\hat{\Gamma}_{\textrm{Virtual}}$ can accordingly be expressed as
\begin{eqnarray}
\hat{\Gamma}_{\textrm{Virtual}}=\hat{\Gamma}_{\textrm{Loop}}+\hat{\Gamma}_{\textrm{CT}}.
\end{eqnarray}

To isolate the ultraviolet (UV) and infrared (IR) divergences, we adopt the dimensional regularization with $D=4-2\epsilon$. The on-mass-shell (OS) scheme is employed to set the renormalization constants for the heavy quark mass ($Z_m$), heavy quark field ($Z_2$), and gluon filed ($Z_3$). The modified minimal-subtraction ($\overline{MS}$) scheme is used for the QCD gauge coupling ($Z_g$). The renormalization constants read $(Q=c,b)$
\begin{eqnarray}
\delta Z_{m}^{OS}&=& -3 C_{F} \frac{\alpha_s}{4\pi}\left[\frac{1}{\epsilon_{\textrm{UV}}}-\gamma_{E}+\textrm{ln}\frac{4 \pi \mu_r^2}{m_{Q}^2}+\frac{4}{3}\right], \nonumber \\
\delta Z_{2}^{OS}&=& - C_{F} \frac{\alpha_s}{4\pi}\left[\frac{1}{\epsilon_{\textrm{UV}}}+\frac{2}{\epsilon_{\textrm{IR}}}-3 \gamma_{E}+3 \textrm{ln}\frac{4 \pi \mu_r^2}{m_{Q}^2}+4\right], \nonumber \\
\delta Z_{3}^{OS}&=& \frac{\alpha_s }{4\pi}\left[(\beta_{0}^{'}-2 C_{A})(\frac{1}{\epsilon_{\textrm{UV}}}-\frac{1}{\epsilon_{\textrm{IR}}})-\frac{4}{3}T_F(\frac{1}{\epsilon_{\textrm{UV}}}-\gamma_E+\textrm{ln}\frac{4\pi\mu_r^2}{m_{c}^2}) \right. \nonumber\\
&& \left. -\frac{4}{3}T_F(\frac{1}{\epsilon_{\textrm{UV}}}-\gamma_E+\textrm{ln}\frac{4\pi\mu_r^2}{m_{b}^2})\right], \nonumber \\
\delta Z_{g}^{\overline{MS}}&=& -\frac{\beta_{0}}{2}\frac{\alpha_s }{4\pi}\left[\frac{1} {\epsilon_{\textrm{UV}}}-\gamma_{E}+\textrm{ln}(4\pi)\right], \label{CT}
\end{eqnarray}
where $\gamma_E$ is the Euler's constant, $\beta_{0}(=\frac{11}{3}C_A-\frac{4}{3}T_Fn_f)$ is the one-loop coefficient of the $\beta$ function, and $\beta_{0}^{'}=\frac{11}{3}C_A-\frac{4}{3}T_Fn_{lf}$. $n_f(=5)$ and $n_{lf}(=n_f-2)$ are the numbers of active quark flavors and light quark flavors, respectively. In ${\rm SU}(3)$, the color factors are given by $T_F=\frac{1}{2}$, $C_F=\frac{4}{3}$, and $C_A=3$.

In calculating $\hat{\Gamma}_{\textrm{Loop}}$, we use FeynArts \cite{Hahn:2000kx} to generate all the involved one-loop diagrams and the corresponding analytical amplitudes; then the package FeynCalc \cite{Mertig:1990an} is applied to tackle the traces of the $\gamma$ and color matrices such that the hard-scattering amplitudes are transformed into expressions with loop integrals. Note that, the $D$-dimension $\gamma$ traces in $\hat{\Gamma}_{\textrm{Loop}}$ involve the $\gamma_{5}$ matrix, and we adopt the following scheme \cite{z decay 15,z decay 17,Korner:1991sx} to deal with it:\\
For Figs. \ref{fig:Virtual}(a), \ref{fig:Virtual}(b), \ref{fig:Virtual}(d), \ref{fig:Virtual}(e), and \ref{fig:Virtual}(f), which contain two $\gamma_{5}$ matrices, we move the two $\gamma_{5}$ together and then obtain an identity matrix by $\gamma_5^2=1$. For the triangle anomalous diagram, i.e. Fig. \ref{fig:Virtual}(c), we choose the same starting point ($Z$-vertex) to write down the amplitudes without the implementation of cyclicity.

In the next step, we utilize our self-written $\textit{Mathematica}$ codes with the implementations of Apart \cite{Feng:2012iq} and FIRE \cite{Smirnov:2008iw} to reduce these loop integrals to a set of irreducible master integrals, which would be numerically evaluated by using the package LoopTools \cite{Hahn:1998yk}. 

\subsubsection{Real corrections}
\begin{figure*}
\includegraphics[width=0.95\textwidth]{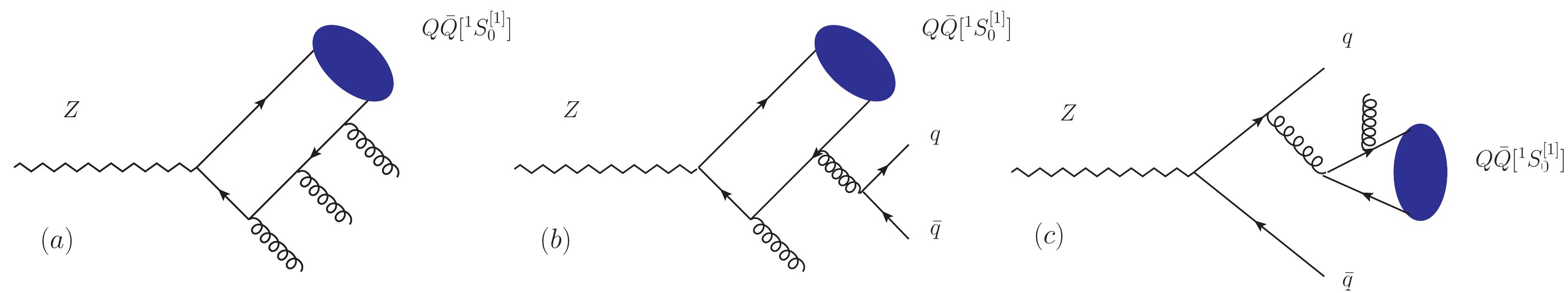}
\caption{\label{fig:Real}
Representative Feynman diagrams of the real corrections to $Z \to Q\bar{Q}[^1S_0^{[1]}]+g+g$ ($Q=c,b$). $``q"$ denotes the light quarks ($u,d,s$).}
\end{figure*}

The real corrections to $Z \to Q\bar{Q}[^1S_0^{[1]}]+g+g$ involve two $1 \to 4$ processes $(q=u,d,s)$,
\begin{eqnarray}
Z & \to & Q\bar{Q}[^1S_0^{[1]}]+g+g+g, \nonumber \\
Z & \to & Q\bar{Q}[^1S_0^{[1]}]+g+q+\bar{q},
\label{pro real}
\end{eqnarray}
whose representative Feynman diagrams are displayed in Fig. \ref{fig:Real}. Note that, in calculating $Z \to Q\bar{Q}[^1S_0^{[1]}]+g+g+g$, we apply the physical polarization tensor, $P_{\mu\nu}$,\footnote{$P_{\mu\nu}=-g_{\mu\nu}+\frac{k_{\mu}\eta_{\mu}+k_{\nu}\eta_{\mu}}{k \cdot \eta}$, where $k$ is the momentum of one of the three final gluons and $\eta$ is conveniently set as the momentum of one of the other two gluons in the final state.} for the polarization summation of the final gluons, thereby avoiding the consideration of the ghost diagrams.

The phase-space integrations of the two processes in Eq. (\ref{pro real}) would generate IR singularities, which can be isolated by slicing the phase space into different regions, namely, the two-cutoff slicing strategy \cite{Harris:2001sx}. By introducing two small cutoff parameters ($\delta_{s}$ and $\delta_{c}$) to decompose the phase space into three parts, $\hat{\Gamma}_{\textrm{Real}}$ can then be written as
\begin{eqnarray}
\hat{\Gamma}_{\textrm{Real}}=\hat{\Gamma}_{\textrm{S}}+\hat{\Gamma}_{\textrm{HC}}+\hat{\Gamma}_{\textrm{H}\overline{\textrm{C}}}.
\end{eqnarray}
$\hat{\Gamma}_{\textrm{S}}$ are the soft terms arsing only from $Z \to Q\bar{Q}[^1S_0^{[1]}]+g+g+g$; $\hat{\Gamma}_{\textrm{HC}}$ denotes the hard-collinear terms, which originate from both the two processes in Eq. (\ref{pro real}). The hard-noncollinear terms $\hat{\Gamma}_{\textrm{H}\overline{\textrm{C}}}$ are finite and we use the FDC package \cite{Wang:2004du} to compute them numerically by means of standard Monte Carlo integration techniques. With the cancellation of the dependences of $\hat{\Gamma}_{\textrm{S}}+\hat{\Gamma}_{\textrm{HC}}$ and $\hat{\Gamma}_{\textrm{H}\overline{\textrm{C}}}$ on $\delta_{s,c}$, the $\hat{\Gamma}_{\textrm{Real}}$ would eventually be independent of the cutoff parameters.

By summing up $\Gamma_{\textrm{Virtual}}$ and $\Gamma_{\textrm{Real}}$, all the divergences involved in the NLO calculations would eventually be canceled, and in the following we will perform the numerical calculations.
\section{Phenomenological results}
Under the approximation of $m_{Q(\bar{Q})}=M_{\eta_Q}/2$ ($Q=c,b$), the quark masses are taken as $m_c=1.5~\textrm{GeV}$ and $m_b=4.7~\textrm{GeV}$ \cite{Zyla:2020zbs}. The other input parameters are set as
\begin{eqnarray}
&&m_Z=91.1876~\textrm{GeV},~~~m_{q/\bar{q}}=0~(q=u,d,s),\nonumber \\
&&\sin^{2}\theta_W=0.226,~~~\alpha=1/128. \label{para}
\end{eqnarray}
To determine $\langle \mathcal O ^{\eta_Q}(^1S_0^{[1]})\rangle$ , we employ the relations to the radial wave functions at the origin,
\begin{eqnarray}
\frac{\langle \mathcal O^{\eta_Q}(^1S_0^{[1]}) \rangle}{2N_c}&=&\frac{1}{4\pi}|R_{\eta_Q}(0)|^2,
\end{eqnarray}
where $|R_{\eta_Q}(0)|^2$ reads \cite{Eichten:1995ch}
\begin{eqnarray}
&&|R_{\eta_c}(0)|^2=0.81~\textrm{GeV}^3, \nonumber \\
&&|R_{\eta_b}(0)|^2=6.477~\textrm{GeV}^3.
\end{eqnarray}

\begin{table*}[htb]
\caption{Decay widths (in units of $\textrm{KeV}$) of $Z \to \eta_c+g+g$  corresponding to different $m_c$ (units: GeV). The superscripts ``$ggg$" and ``$gq\bar{q}$" stand for $Z \to c\bar{c}[^1S_0^{[1]}]+g+g+g$ and $Z \to c\bar{c}[^1S_0^{[1]}]+g+q+\bar{q}$, respectively, ``v(av)" for the (axial-)vector part, and ``frag" for the processes in Fig. \ref{fig:Real}(c). $K$ is identical to $\Gamma_{\textrm{NLO}}/\Gamma_{\textrm{LO}}$. The cutoff parameters are taken as $\delta_{s}=1 \times 10^{-3}$ and $\delta_{c}=2 \times 10^{-5}$.}
\label{etac decay width}
\begin{tabular}{cccccccccccc}
\hline\hline
$~\mu_r~$ & $~m_c~$ & $~\alpha_s~$ & $~~\Gamma_{\textrm{LO}}~~$ & $\Gamma_{\textrm{Vir+S+HC}}$ & $~~\Gamma^{ggg}_{\textrm{H}\overline{\textrm{C}}}~~$ & $~~\Gamma^{gq\bar{q}_{\textrm{av}}}_{\textrm{H}\overline{\textrm{C}}}~~$ & $~~\Gamma^{gq\bar{q}_{\textrm{v}}}_{\textrm{H}\overline{\textrm{C}}}~~$ & $~~\Gamma_{\textrm{NLO}}~~$ & $~~K~~$ & $~~\Gamma^{gq\bar{q}}_{\textrm{frag}}~~$\\ \hline
$~$ & $1.4$ & $0.26573$ & $5.721$ & $-110.2$ & $104.8$ & $69.54$ & $25.04$ & $94.89$ & $16.6$ & $91.31$\\
$2m_c$ & $1.5$ & $0.25864$ & $4.828$  & $-90.29$ & $85.97$ & $49.22$ & $17.58$ & $67.31$ & $13.9$ & $64.08$\\
$~$ & $1.6$ & $0.25235$ & $4.123$ & $-75.01$ & $71.43$ & $35.70$ & $12.64$ & $48.88$ & $11.9$ & $46.07$\\ \hline
$~$ & $1.4$ & $0.11916$ & $1.150$ & $-8.772$ & $9.455$ & $6.270$ & $2.258$ & $10.36$ & $9.01$ & $8.233$\\
$m_Z$ & $1.5$ & $0.11916$ & $1.025$ & $-7.812$ & $8.401$ & $4.814$ & $1.719$ & $8.147$ & $7.95$ & $6.270$\\
$~$ & $1.6$ & $0.11916$ & $0.919$ & $-7.002$ & $7.521$ & $3.759$ & $1.330$ & $6.527$ & $7.10$ & $4.851$\\ \hline\hline
\end{tabular}
\end{table*}

\begin{table*}[htb]
\caption{Decay widths (in units of $\textrm{KeV}$) of $Z \to \eta_b+g+g$  corresponding to different $m_b$ (unit: GeV). The superscripts ``$ggg$" and ``$gq\bar{q}$" stand for $Z \to b\bar{b}[^1S_0^{[1]}]+g+g+g$ and $Z \to b\bar{b}[^1S_0^{[1]}]+g+q+\bar{q}$, respectively, ``v(av)" for the (axial-)vector part, and ``frag" for the processes in Fig. \ref{fig:Real}(c). $K$ is identical to $\Gamma_{\textrm{NLO}}/\Gamma_{\textrm{LO}}$. The cutoff parameters are taken as $\delta_{s}=1 \times 10^{-3}$ and $\delta_{c}=2 \times 10^{-5}$.}
\label{etab decay width}
\begin{tabular}{cccccccccccc}
\hline\hline
$~\mu_r~$ & $~m_b~$ & $~\alpha_s~$ & $~~\Gamma_{\textrm{LO}}~~$ & $\Gamma_{\textrm{Vir+S+HC}}$ & $~~\Gamma^{ggg}_{\textrm{H}\overline{\textrm{C}}}~~$ & $~~\Gamma^{gq\bar{q}_{\textrm{av}}}_{\textrm{H}\overline{\textrm{C}}}~~$ & $~~\Gamma^{gq\bar{q}_{\textrm{v}}}_{\textrm{H}\overline{\textrm{C}}}~~$ & $~~\Gamma_{\textrm{NLO}}~~$ & $~~K~~$ & $~~\Gamma^{gq\bar{q}}_{\textrm{frag}}~~$\\ \hline
$~$ & $4.6$ & $0.18422$ & $2.515$ & $-31.35$ & $29.94$ & $2.192$ & $0.420$ & $3.717$ & $1.48$ & $1.533$\\
$2m_b$ & $4.7$ & $0.18326$ & $2.383$  & $-29.49$ & $28.17$ & $2.007$ & $0.374$ & $3.444$ & $1.44$ & $1.363$\\
$~$ & $4.8$ & $0.18234$ & $2.260$ & $-27.77$ & $26.52$ & $1.843$ & $0.333$ & $3.186$ & $1.41$ & $1.215$\\ \hline
$~$ & $4.6$ & $0.11916$ & $1.052$  & $-7.783$ & $8.103$ & $0.593$ & $0.114$ & $2.079$ & $1.98$ & $0.415$\\
$m_Z$ & $4.7$ & $0.11916$ & $1.007$  & $-7.440$ & $7.742$ & $0.552$ & $0.103$ & $1.964$ & $1.95$ & $0.374$\\
$~$ & $4.8$ & $0.11916$ & $0.965$ & $-7.117$ & $7.402$ & $0.514$ & $0.093$ & $1.857$ & $1.92$ & $0.339$\\ \hline\hline
\end{tabular}
\end{table*}

\begin{figure*}
\includegraphics[width=0.495\textwidth]{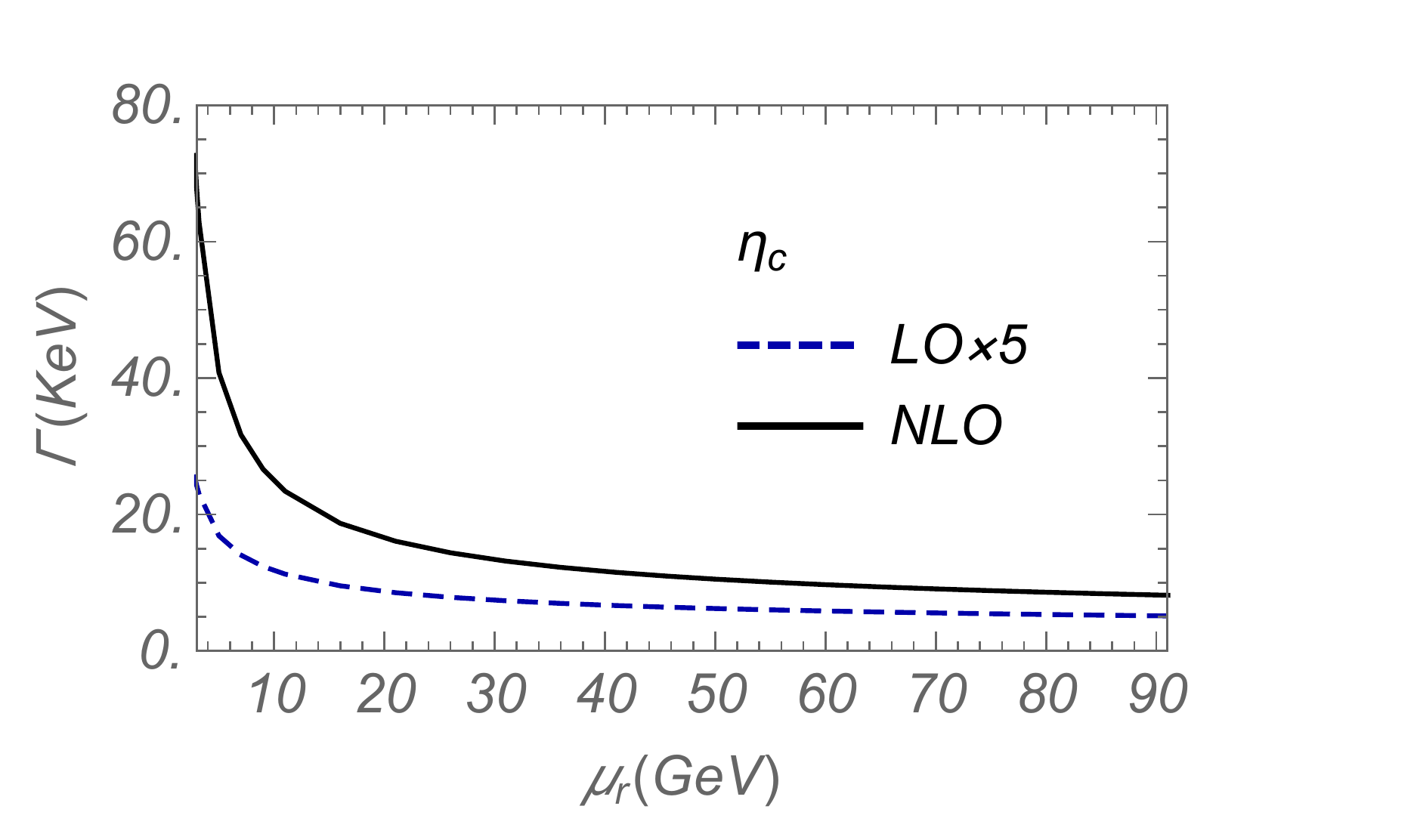}
\includegraphics[width=0.495\textwidth]{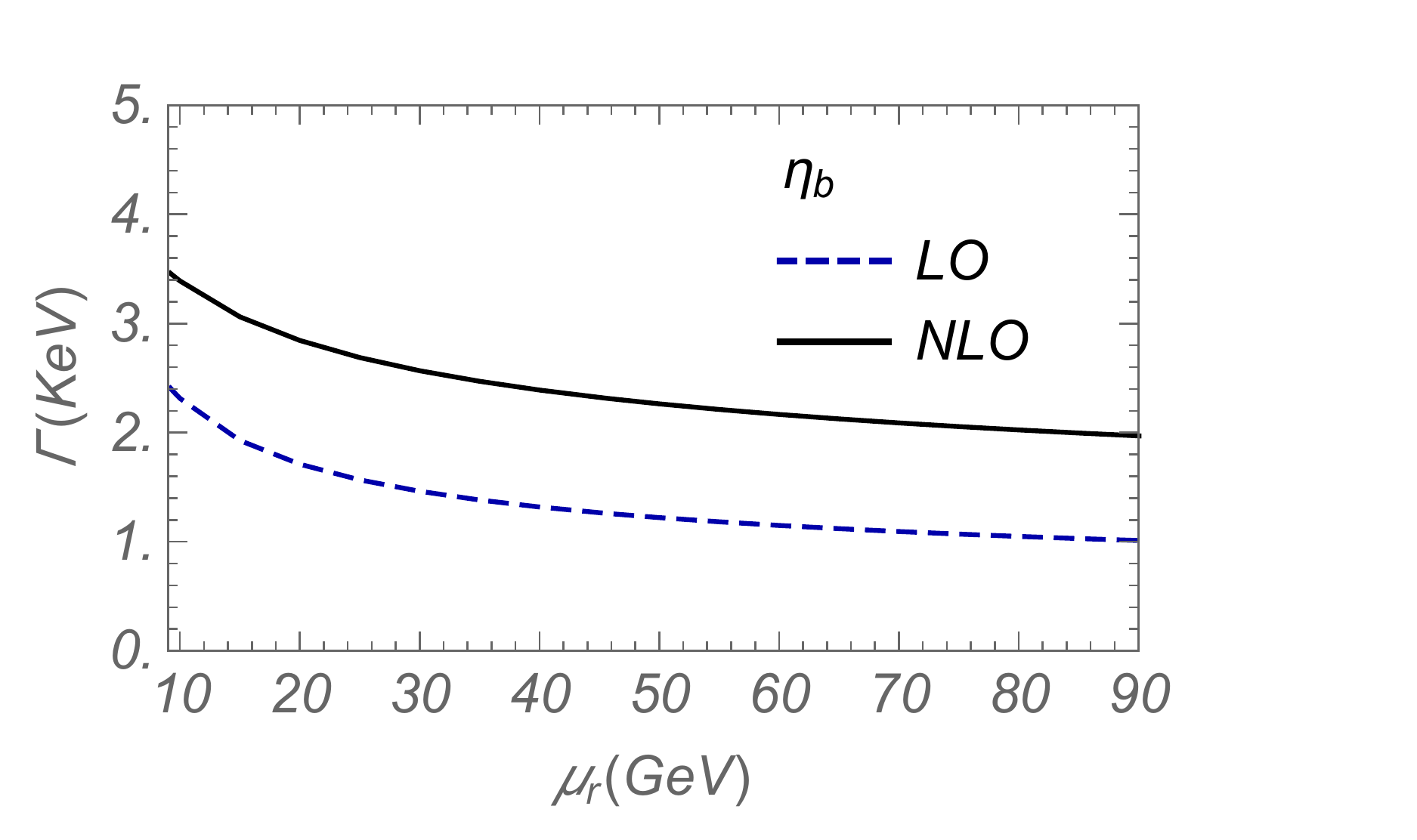}
\caption{\label{fig:mur}
Decay widths of $Z \to \eta_Q+g+g$ ($Q=c,b$) as a function of the renormalization scale $\mu_r$. $m_c=1.5$ GeV and $m_b=4.7$ GeV.}
\end{figure*}

We summarize the predicted decay widths of $Z \to \eta_Q+g+g$ in Tables. \ref{etac decay width} and \ref{etab decay width}. Inspecting the two tables, one can observe
\begin{itemize}
\item[i)]
For $Z \to \eta_c+g+g$, $\Gamma_{\textrm{Vir+S+HC}}$ severely cancels the large contribution of $\Gamma^{ggg}_{\textrm{H}\overline{\textrm{C}}}$; the other part in $\Gamma_{\textrm{H}\overline{\textrm{C}}}$, i.e. $\Gamma^{gq\bar{q}}_{\textrm{H}\overline{\textrm{C}}}$, which is dominated by the significant contributions of the gluon-fragmentation structures [Fig. \ref{fig:Real}(c); cf. $\Gamma^{gq\bar{q}}_{\textrm{frag}}$ in Table \ref{etac decay width}], is comparable with $\Gamma^{ggg}_{\textrm{H}\overline{\textrm{C}}}$ and then enhances the LO results to an extremely large extent, as pictorially shown in the left panel of Fig. \ref{fig:mur}. In other words, the large $K$ factors in Table I can mainly be attributed to the contributions of Fig. \ref{fig:Real}(c), which is gauge invariant and free of divergences. $\Gamma_{\textrm{NLO}}$ appears to be more sensitive than $\Gamma_{\textrm{LO}}$ on the choice of the $c$-quark mass, which can be understood by the fact that the dominant gluon-fragmentation contributions in $\Gamma^{gq\bar{q}}_{\textrm{H}\overline{\textrm{C}}}$ depend heavily on the value of $m_c$. 
\item[ii)]
As for $\eta_b$, there still holds a severe cancellation between $\Gamma_{\textrm{Vir+S+HC}}$ and $\Gamma^{ggg}_{\textrm{H}\overline{\textrm{C}}}$; however, since the impact of the gluon-fragmentation structure, $g^{*} \to \eta_b +g$, is greatly weakened by the large mass of $\eta_b$ (cf. $\Gamma^{gq\bar{q}}_{\textrm{frag}}$ in Table \ref{etab decay width}), $\Gamma^{gq\bar{q}}_{\textrm{H}\overline{\textrm{C}}}$ contributes just slightly. As a result, the QCD corrections to $Z \to \eta_b+g+g$ appear to be much wilder than the $\eta_c$ case, which can clearly be seen by the second panel in Fig. \ref{fig:mur}.
\end{itemize}

Now, we compare the contributions of $Z \to \eta_Q+g+g$ ($Q=c,b$) with those of $Z \to \eta_Q+Q+\bar{Q}$. Taking $\mu_r=2m_{c,b}$ with $m_c=1.5$ GeV and $m_b=4.7$ GeV, we have
\begin{eqnarray}
\Gamma^{c\bar{c}}_{\textrm{LO}}&=&99.90~\textrm{KeV}, \nonumber \\
\Gamma^{b\bar{b}}_{\textrm{LO}}&=&12.23~\textrm{KeV},
\label{QQ results}
\end{eqnarray}
and then
\begin{eqnarray}
\frac{\Gamma^{gg}_{\textrm{LO}}}{\Gamma^{c\bar{c}}_{\textrm{LO}}}&=&4.83\%,~~~\frac{\Gamma^{gg}_{\textrm{NLO}}}{\Gamma^{c\bar{c}}_{\textrm{LO}}}=67.4\%, \nonumber \\
\frac{\Gamma^{gg}_{\textrm{LO}}}{\Gamma^{b\bar{b}}_{\textrm{LO}}}&=&19.5\%,~~~\frac{\Gamma^{gg}_{\textrm{NLO}}}{\Gamma^{b\bar{b}}_{\textrm{LO}}}=28.1\%, \label{ratio}
\end{eqnarray}
where ``$gg$" stands for $Z \to \eta_Q+g+g$, and ``$Q\bar{Q}$" stands for $Z \to \eta_Q+Q+\bar{Q}$. One can find, after including the newly calculated QCD corrections to $Z \to \eta_Q+g+g$, that the $gg$ process would be comparable with the $Q\bar{Q}$ one.

\begin{figure*}
\includegraphics[width=0.495\textwidth]{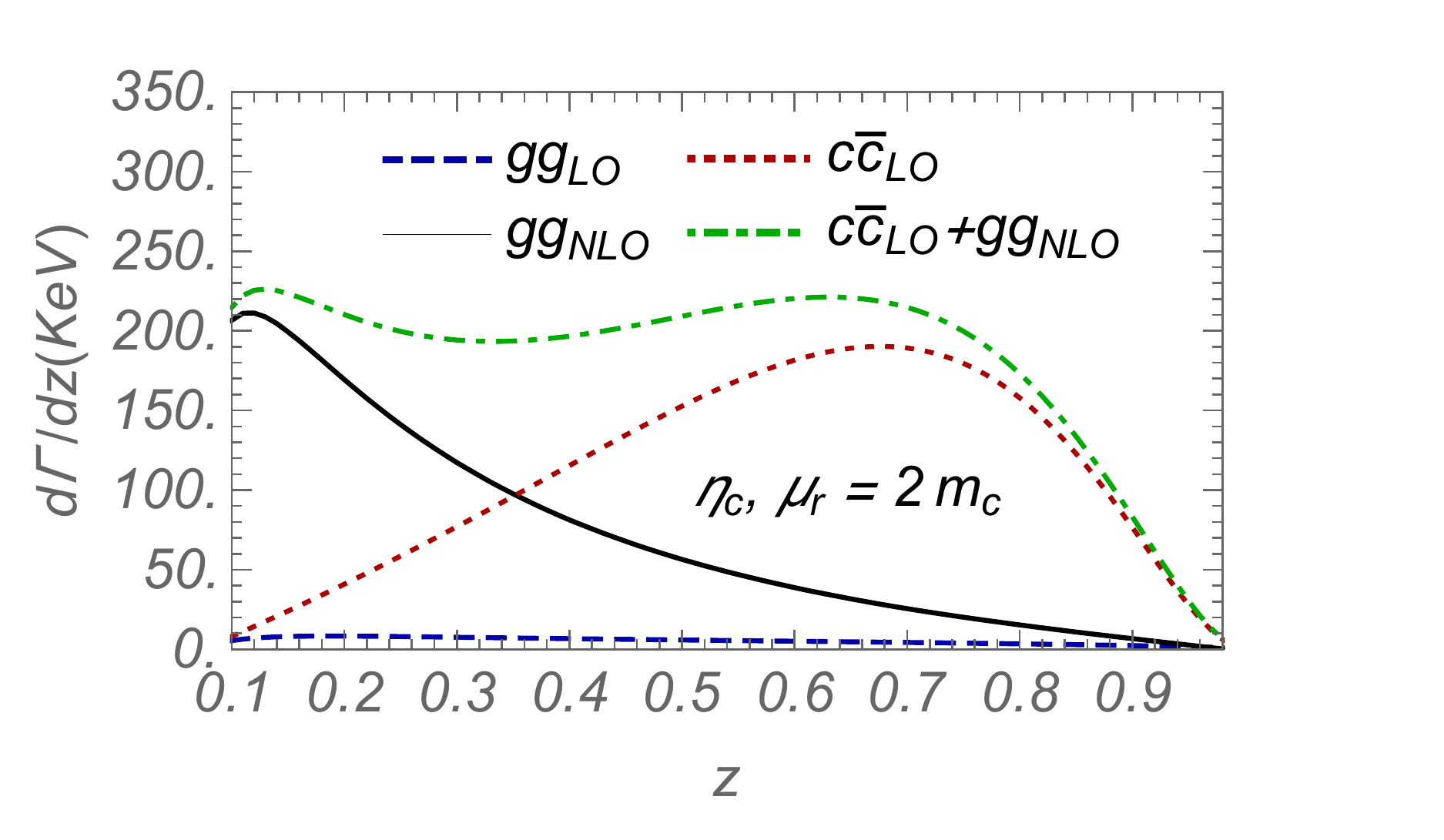}
\includegraphics[width=0.495\textwidth]{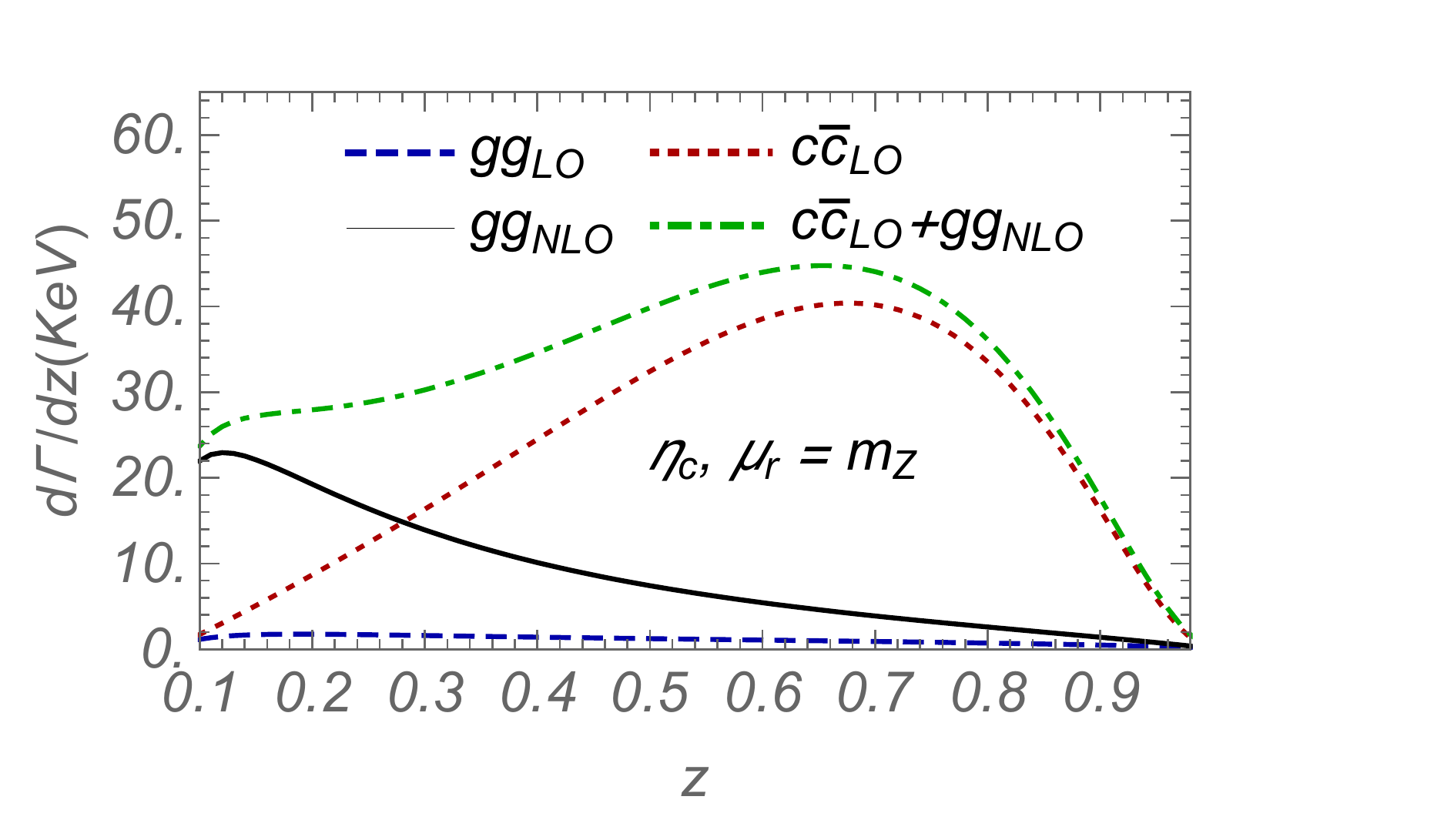}
\includegraphics[width=0.495\textwidth]{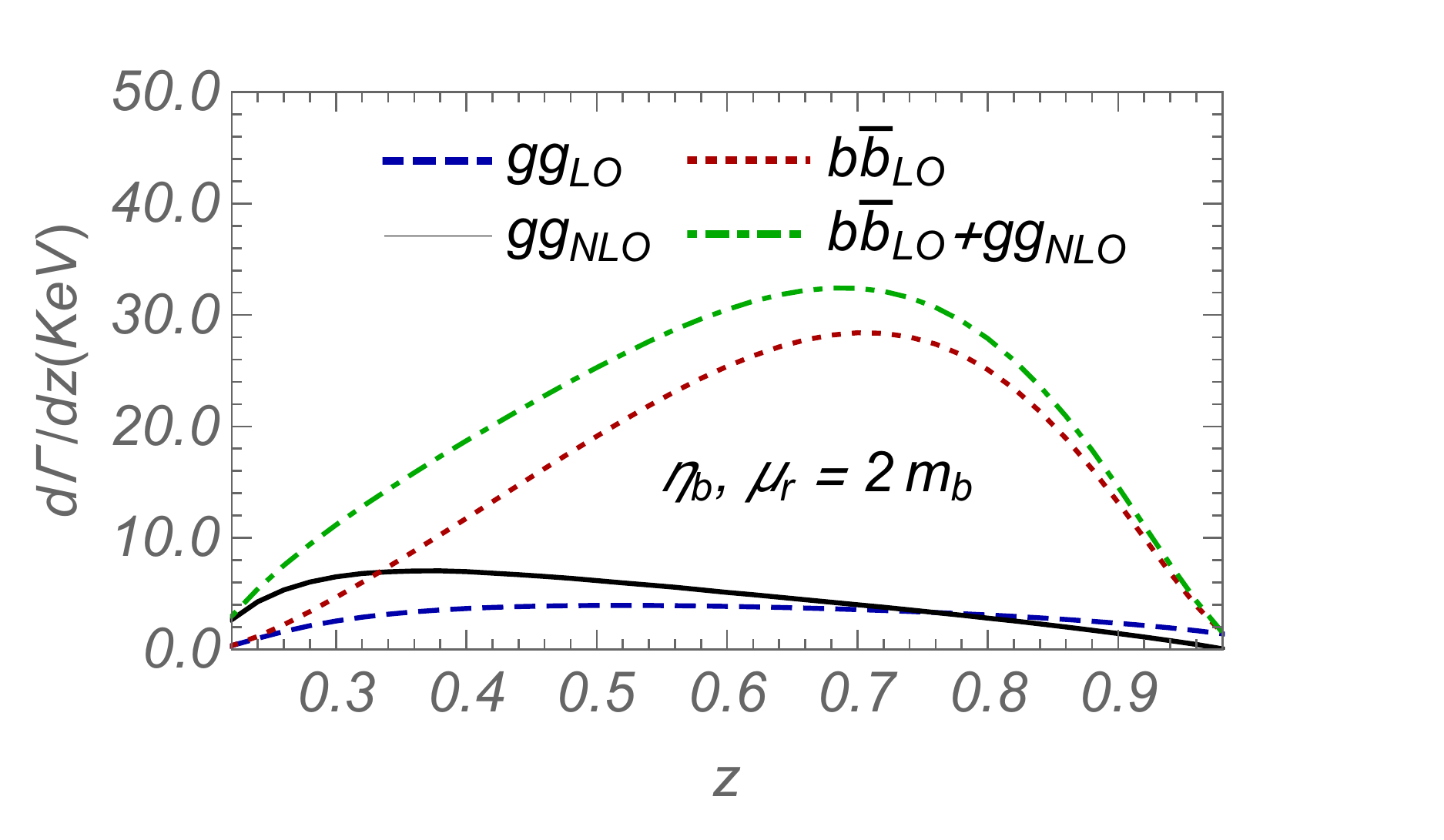}
\includegraphics[width=0.495\textwidth]{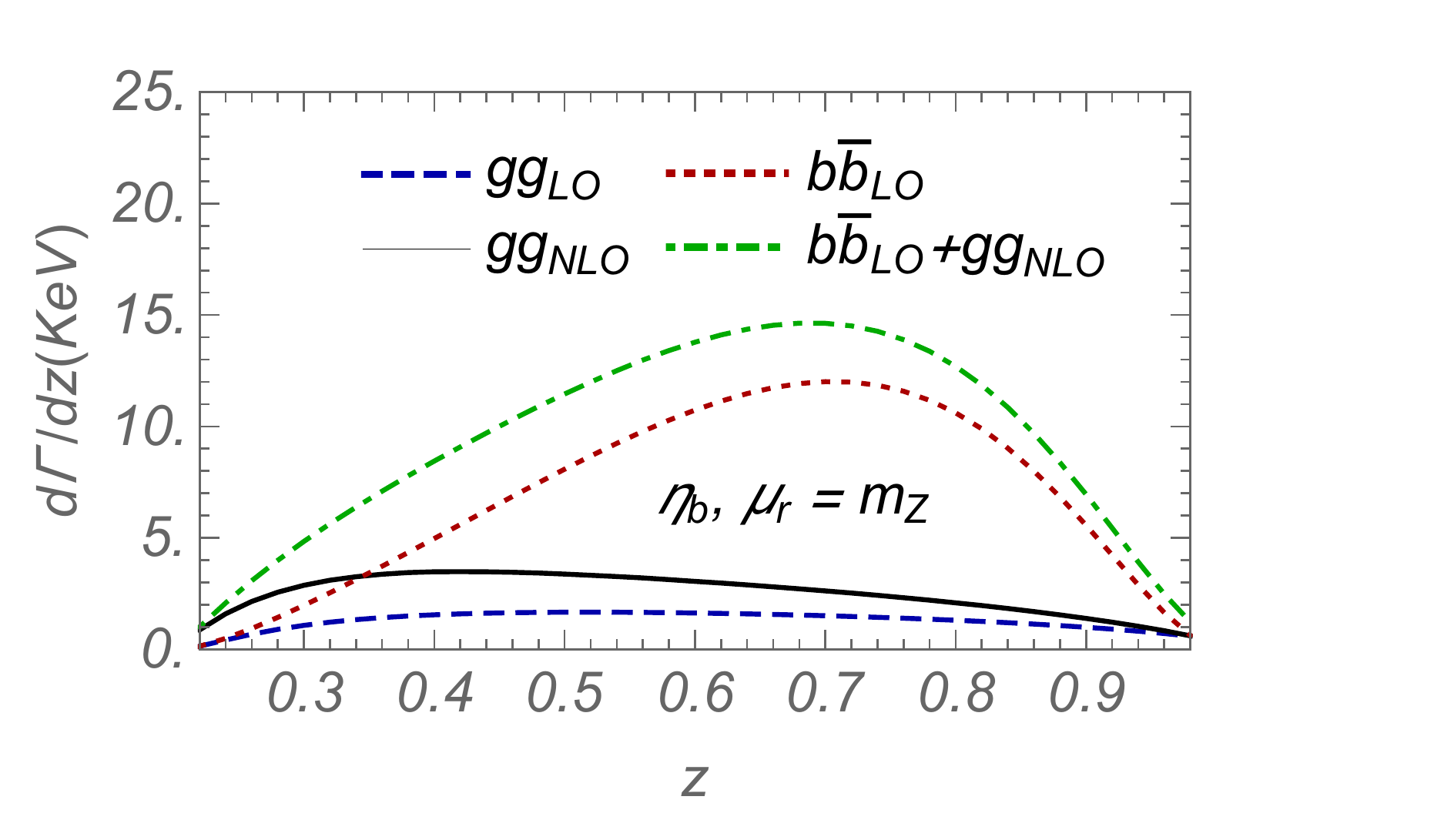}
\caption{\label{fig:zdis}
$\eta_Q$ ($Q=c,b$) energy distributions with $z$ defined as $\frac{2E_{\eta_Q}}{m_Z}$; $``gg(Q\bar{Q})"$ denotes the process of $Z \to \eta_Q+gg(Q\bar{Q})$. $m_c=1.5$ GeV and $m_b=4.7$ GeV.}
\end{figure*}

In Fig. \ref{fig:zdis}, the $\eta_Q$ energy distributions are drawn with $z$ defined as $\frac{2E_{\eta_Q}}{m_Z}$. It can be seen that,
\begin{itemize}
\item[i)]
The dominant contributions in $\Gamma^{\textrm{LO}}_{Z \to \eta_c+c+\bar{c}}$ arise from the region of $z \simeq 0.7$, while the peak of $\frac{d\Gamma^{\textrm{LO}}_{Z \to \eta_c+g+g}}{dz}$ lies in the vicinity of $z \simeq 0.2$. By incorporating the QCD corrections, the $gg$ results are notably enhanced, especially at the small- and mid-$z$ regions. As a result, adding the $gg$ contributions would greatly increase the differential decay widths given by $Z \to \eta_c+c+\bar{c}$, which can clearly be seen by the huge discrepancy between the two lines referring to $c\bar{c}_{\textrm{LO}}$ with or without $gg_{\textrm{NLO}}$ in the two upper panels of Fig. \ref{fig:zdis}.
\item[ii)]
Regarding $\eta_b$, there also exists an evident peak of $\frac{d\Gamma^{\textrm{LO}}_{Z \to \eta_b+b+\bar{b}}}{dz}$ around $z \simeq 0.7$; in $Z \to \eta_b+g+g$ at LO, the mid-$z$ regions ($z \simeq 0.5$) contribute dominantly. With the QCD corrections, the $gg$ process would evidently raise the lines given by $Z \to \eta_b+b+\bar{b}$, as manifested by the large difference in height of the line of $b\bar{b}_{\textrm{LO}}$ and that of $b\bar{b}_{\textrm{LO}}+gg_{\textrm{NLO}}$ in the two lower panels of Fig. \ref{fig:zdis}.
\end{itemize}

To summarize, our newly calculated QCD corrections to $Z \to \eta_{Q}[^1S_0^{[1]}]+g+g$ could enormously enhance its LO results, and then greatly elevate the phenomenological significance of the $gg$ process in $Z$ decaying into inclusive $\eta_c$.

Inspired by the large contributions of Fig. \ref{fig:Real}(c), at last, we investigate the significance of $Z \to c\bar{c}[^1S_0^{[1]}]+g+b+\bar{b}$ and $Z \to b\bar{b}[^1S_0^{[1]}]+g+c+\bar{c}$,\footnote{The processes of $Z \to c\bar{c}[^1S_0^{[1]}]+g+c+\bar{c}$ and $Z \to b\bar{b}[^1S_0^{[1]}]+g+b+\bar{b}$, which include IR singularities, should be categorized as parts of the real corrections to $Z \to c\bar{c}[^1S_0^{[1]}]+c+\bar{c}$ and $Z \to b\bar{b}[^1S_0^{[1]}]+b+\bar{b}$, respectively.} which also involve the gluon-fragmentation structures. The two processes are free of divergences, and by straightforward calculations under $\mu_r=2m_{c,b}$ ($m_c=1.5$ GeV and $m_b=4.7$ GeV), we have
\begin{eqnarray}
\Gamma_{Z \to c\bar{c}[^1S_0^{[1]}]+g+b+\bar{b}}&=&20.01~\textrm{KeV}, \nonumber \\
\Gamma_{Z \to b\bar{b}[^1S_0^{[1]}]+g+c+\bar{c}}&=&0.547~\textrm{KeV}.\label{QQfrag}
\end{eqnarray}
As compared to Eq. (\ref{QQ results}), the above two processes are indispensable for the inclusive $\eta_{c,b}$ yield in $Z$-boson decay.

\section{Summary}
In this manuscript, we achieve the first NLO corrections to $Z \to \eta_Q+g+g~(Q=c,b)$ through the CS state of $Q\bar{Q}[^1S_0^{[1]}]$. We find that the newly calculated QCD corrections can noticeably enhance its LO results, following which the $gg$ process would contribute comparably to the CS-dominant process $Z \to \eta_Q[^1S_0^{[1]}]+Q+\bar{Q}$. Moreover, with the QCD corrections, the $gg$ process would profoundly influence the existing CS-predicted $\eta_Q$ energy distribution. Therefore, to arrive at a strict CS prediction of $Z \to \eta_Q+X$, besides $Z \to \eta_Q[^1S_0^{[1]}]+Q+\bar{Q}$, it appears mandatory to take $Z \to \eta_Q[^1S_0^{[1]}]+g+g$ into consideration as well.
\section{Acknowledgments}
\noindent{\bf Acknowledgments}:
This work is supported in part by the Natural Science Foundation of China under the Grants No. 12065006, and by the Project of GuiZhou Provincial Department of Science and Technology under Grant No. QKHJC[2019]1167. and No. QKHJC[2020]1Y035.\\

\end{document}